\documentclass[epj]{svjour}
\usepackage{epsfig}
\usepackage{bm}
\begin{document}
\title{Two-pion-exchange parity-violating 
potential and $\vec{n} p \rightarrow d \gamma$}
\author{
C. H. Hyun \inst{1} \and 
S. Ando \inst{1} \and
B. Desplanques \inst{2}
}                     
\institute{Department of Physics and Institute of Basic Science,
Sungkyunkwan University, Suwon 440-746, Korea \and 
Laboratoire de Physique Subatomique et de Cosmologie 
(UMR CNRS/IN2P3-UJF-INPG), F-38026 Grenoble Cedex, France}
\date{Received: date / Revised version: date}
\abstract{
We calculate the parity-violating nucleon-nucleon potential
in heavy-baryon chiral perturbation theory up to the
next-to-next-to-leading order.
The one-pion exchange comes in the leading order
and the next-to-next-to-leading order consists of
two-pion-exchange and the two-nucleon contact terms.
In order to investigate the effect of the higher order
contributions, we calculate the parity-violating asymmetry 
in $\vec{n} p \rightarrow d \gamma$ at the threshold.
The one-pion dominates the physical observable and
the two-pion contribution is about or less than 10\% of the 
one-pion contribution.
\PACS{
      {21.30.Fe}{Forces in hadronic systems and effective interactions}\and
      {12.15.Ji}{Applications of electroweak models to specific processes}
     } 
} 
\maketitle
\section{Introduction}
\label{intro}
We employ the idea of effective field theory (EFT) 
to derive the parity-violating (PV) nucleon-nucleon ($NN$) potential 
and apply it to the calculation of the PV photon asymmetry $A_\gamma$ in
$\vec{n}p \rightarrow d\gamma$ at threshold.
The PV $NN$ potential is obtained by replacing a
parity-conserving (PC) vertex in the strong $NN$
potential with a PV vertex.
Most of the low energy PV calculations have been
relying on one-meson exchange potential, the so-called
DDH potential~\cite{ddh80}.
Some literature investigated the PV two-pion-exchange potential
(TPEP) in the past~\cite{bd_plb72,pr_plb73,cd_npb74},
and it was revived quite recently in the light
of EFT~\cite{zhu_npa05}.
In ref.~\cite{zhu_npa05}, a thorough derivation of the
PV potential is performed up to the 
next-to-next-to-leading order (NNLO)
in both pionless and pionful EFTs.
Since the PV asymmetry in $\vec{n}p \rightarrow d\gamma$
is sensitive to the pion exchange contribution,
we adopt 
heavy-baryon chiral perturbation (HB$\chi$PT)
and obtain the potential relevant to
$\vec{n}p \rightarrow d\gamma$ up to NNLO.

Theoretical estimations of $A_\gamma$ have been 
extensively worked out with the DDH potential
\cite{bd_plb2001,hyun_plb2001,rocco_prc2004,hyun_epja05}.
The results with various strong interaction models turn
out basically identical. $A_\gamma$ is dominated
by the PV one-pion-exchange potential (OPEP) and 
the heavy-meson contribution is negligible. 
Thus it is discussed that the measurements of $A_\gamma$
could provide us an opportunity to determine
the weak pion-nucleon coupling constant $h_\pi^1$.
On the other hand, possibility of a 10\% effect
from the PV TPEP has been discussed~\cite{bd_pr98}. 

In this work we present the first estimation of
the intermediate-range contribution to $A_\gamma$.
We employ the Argonne v18 potential for the PC potential 
and the Siegert's theorem for the current operators.
The PV potential relevant to $\vec{n}p \rightarrow d\gamma$ in our
study comprises OPEP and TPEP.
There are various terms other than the TPEP that appear at the
NNLO: two-nucleon contact term, higher order correction of PC
and PV $\pi NN$ vertices, and etc. In the present work,
we retain only the TPEP and neglect the remaining NNLO contributions
for simplicity. The effect of the remaining terms will be considered
elsewhere \cite{had_prepare}.

The NNLO calculation will allow us to estimate the order
and the magnitude of higher order corrections, which will be
important in pinning down the value of $h^1_\pi$ and its
uncertainty.
At the same time, the NNLO contribution will provide a criterion
for the validity of EFT approach to the PV phenomena.
\section{Formalism}
\label{sec:formalism}

\subsection{PV potential}
\label{sec:potential}

\begin{figure*}
\begin{center}
\resizebox{0.6\textwidth}{!}{%
  \includegraphics{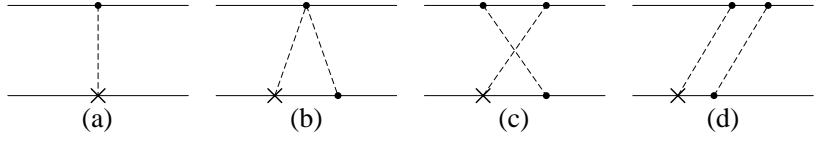}
}
\end{center}
\caption{Diagrams for PV OPE and TPE potentials.
The diagram (a) is for the OPE term
and the diagram (b-d) are for the TPE ones .
Lines (dashed lines) denote nucleons (pions),
vertices with a dot represent PC vertices,
vertices with ``$\bm{\times}$" represent the PV 
vertex proportional to $h^1_\pi$.}
\label{fig:pep}
\end{figure*}

The PV potential relevant to our estimation of $A_\gamma$ in
momentum space has the form
\begin{eqnarray}
\tilde{V}_i(\bm{q}) = i (\bm{\tau}_1 \times \bm{\tau}_2)^z
(\bm{\sigma}_1 + \bm{\sigma}_2) \cdot \bm{q} \,\,
\tilde{v}_i (q),
\label{eq:vq}
\end{eqnarray}
where $q \equiv |\bm{q}|$ and $\bm{q} = \bm{p}_1 - \bm{p}_2$.
OPE and TPE terms are obtained as
\begin{eqnarray}
\tilde{v}_{1\pi}(q) &=& - \frac{g_A h^1_\pi}{2 \sqrt{2} f_\pi}
\frac{1}{q^2 + m_\pi^2},
\label{eq:vq_ope} \\
\tilde{v}_{2\pi}(q) &=& \sqrt{2}\pi \frac{h^1_\pi}{\Lambda^3_\chi}
\left\{ g_A \tilde{L}(q)
- g^3_A \left[3\tilde{L}(q) - \tilde{H}(q) \right]
\right\},
\label{eq:vq_tpe}
\end{eqnarray}
with
\begin{eqnarray}
\tilde{L}(q) &=& \frac{\sqrt{q^2 + 4 m_\pi^2}}{q}
\ln \left( \frac{\sqrt{q^2+4 m_\pi^2} + q}{2 m_\pi} \right),
\label{eq:lq} \\
\tilde{H}(q) &=& \frac{4 m_\pi^2}{q^2 + m_\pi^2} \tilde{L}(q)\, ,
\label{eq:hq}
\end{eqnarray}
where $g_A$ is the axial coupling constant, $f_\pi$
the pion decay constant and $\Lambda_\chi=4\pi f_\pi$.
The potential of eq.~(\ref{eq:vq}) in the coordinate space can
be written as
\begin{eqnarray}
V_i(\bm{r}) &=& \int \frac{d^3 \bm{q}}{(2 \pi)^3}
\tilde{V}_i(\bm{q})\, {\rm e}^{-i \bm{q}\cdot \bm{r}} \nonumber \\
&=& i (\bm{\tau}_1 \times \bm{\tau}_2)^z
(\bm{\sigma}_1 + \bm{\sigma}_2) \cdot \left[ \bm{p},\, v_i(r) \right],
\label{eq:Vv}
\end{eqnarray}
where $\bm{p}$ is the conjugate momentum of the relative coordinate
$\bm{r} \equiv \bm{r}_1 - \bm{r}_2$.
For easier numerical calculation, we cast eqs.~(\ref{eq:lq},\ref{eq:hq})
into the dispersion relations as
\begin{eqnarray}
\tilde{L}(q) &=& - \int^{\infty}_{4m^2_\pi}
\frac{dt'}{2 \sqrt{t'}} \sqrt{t'\! -\! 4m^2_\pi} \left(
\frac{1}{t'\!+\!q^2}\! - \!\frac{1}{t'\! -\! 4m^2_\pi} \right), \\
\tilde{H}(q) &=& \frac{4m^2_\pi}{2} \int^{\infty}_{4m^2_\pi}
\frac{dt'}{\sqrt{t'}}\frac{1}{\sqrt{t'-4m^2_\pi}}
\frac{1}{t'+q^2}.
\end{eqnarray}
Furthermore, we introduce a monopole form factor of the type
$\Lambda^2/(\Lambda^2 + q^2)$ in the Fourier transformation
of eq.~(\ref{eq:vq}).
The roles of the form factor and the cutoff are
(i) to make the numerical calculation more easier and efficient,
and (ii) to cut away high momentum region where the dynamics
is essentially unknown and irrelevant to the low energy processes.
With the form factor, we rewrite the potential in coordinate
space as
\begin{eqnarray}
V^\Lambda_i(\bm{r}) = i (\bm{\tau}_1 \times \bm{\tau}_2)^z
(\bm{\sigma}_1 + \bm{\sigma}_2) \cdot
[ \bm{p},\, v^\Lambda_i(r)]
\end{eqnarray}
where
\begin{eqnarray}
v^\Lambda_{1\pi} &=& \frac{g_A h^1_\pi}{2 \sqrt{2} f_\pi}
\frac{\Lambda^2}{\Lambda^2 - m^2_\pi} \frac{1}{4\pi r}
({\rm e}^{-m_\pi r} - {\rm e}^{-\Lambda r}), 
\label{eq:vr_ope}\\
v^\Lambda_{2\pi} &=& \sqrt{2}\pi \frac{h^1_\pi}{\Lambda^3_\chi}
\left\{ g_A L^\Lambda (r)
- g^3_A \left[3 L^\Lambda (r) - H^\Lambda (r) \right]
\right\},
\label{eq:vr_tpe}
\end{eqnarray}
with
\begin{eqnarray}
L^\Lambda (r) &=& \frac{\Lambda^2}{8\pi r}
\int^\infty_{4 m^2_\pi} \frac{dt'}{\sqrt{t'}}\sqrt{t'-4m^2_\pi} 
\nonumber \\
&\times&
\left(\frac{{\rm e}^{-\sqrt{t'} r} - {\rm e}^{-\Lambda r}}{\Lambda^2 - t'} 
- \frac{{\rm e}^{-\Lambda r}}{t' - 4m^2_\pi}\right), \\
H^\Lambda (r) &=& \frac{m^2_\pi \Lambda^2}{2 \pi r}
\int^\infty_{4 m^2_\pi} \frac{dt'}{\sqrt{t'}}\frac{1}{\sqrt{t'\!-\!4m^2_\pi}}
\frac{{\rm e}^{-\sqrt{t'} r}\! -\! {\rm e}^{-\Lambda r}}{\Lambda^2\! -\! t'}.
\end{eqnarray}

\subsection{Asymmetry in $\vec{n}p\rightarrow d\gamma$}
\label{sec:asymmetry}

The photon asymmetry in $\vec{n} p \rightarrow d \gamma$,
$A_\gamma$, is defined from the differential cross section
of the process as
\begin{eqnarray}
\frac{d \sigma}{d \Omega} \propto 1 + A_\gamma \cos\theta,
\end{eqnarray}
where $\theta$ is the angle between the neutron polarization and
the out-going photon momentum.
Non-zero $A_\gamma$ values arise from the interference of opposite
parity transition amplitudes, {\it e.g.} M1 and E1.

At the thermal energy where the process occurs,
lowest order EM operators may suffice, therefore
we consider the 
E1 operator,
\begin{eqnarray}
\bm{J}_{\rm E1} &=& -i \frac{\omega_\gamma}{4}\,
(\tau^z_1 - \tau^z_2)\, \bm{r} \, ,
\label{eq:op_e1}
\end{eqnarray}
where 
$\omega_\gamma$ is 
the energy of out-going photon.
At the leading order of $h^1_\pi$,
$A_\gamma$ is proportional to
$h^1_\pi$, and we can write $A_\gamma$ as
\begin{equation}
A_\gamma = a_\gamma h^1_\pi,
\label{eq:agamma}
\end{equation}
with
\begin{eqnarray}
a_\gamma =
- 2 \frac{{\rm Re}\left({\cal M}_1 {\cal E}_1^*\right)}{|{\cal M}_1|^2},
\end{eqnarray}
where ${\cal E}_1$ and ${\cal M}_1$ are matrix elements of the E1 and M1
transitions, respectively.
Analytic forms of these amplitudes can be found in \cite{hyun_epja05}.

\section{Result}

\begin{figure*}
\begin{center}
\epsfig{file = 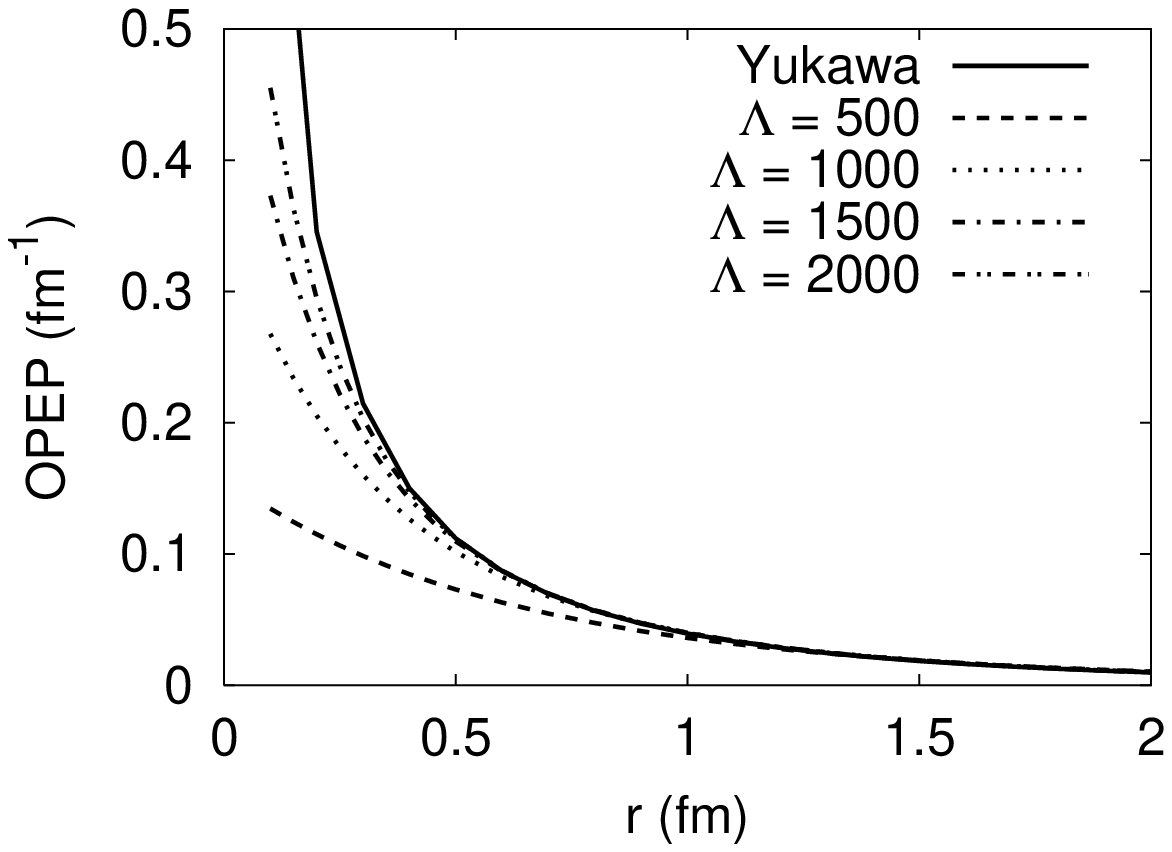, width=5cm}
\epsfig{file = 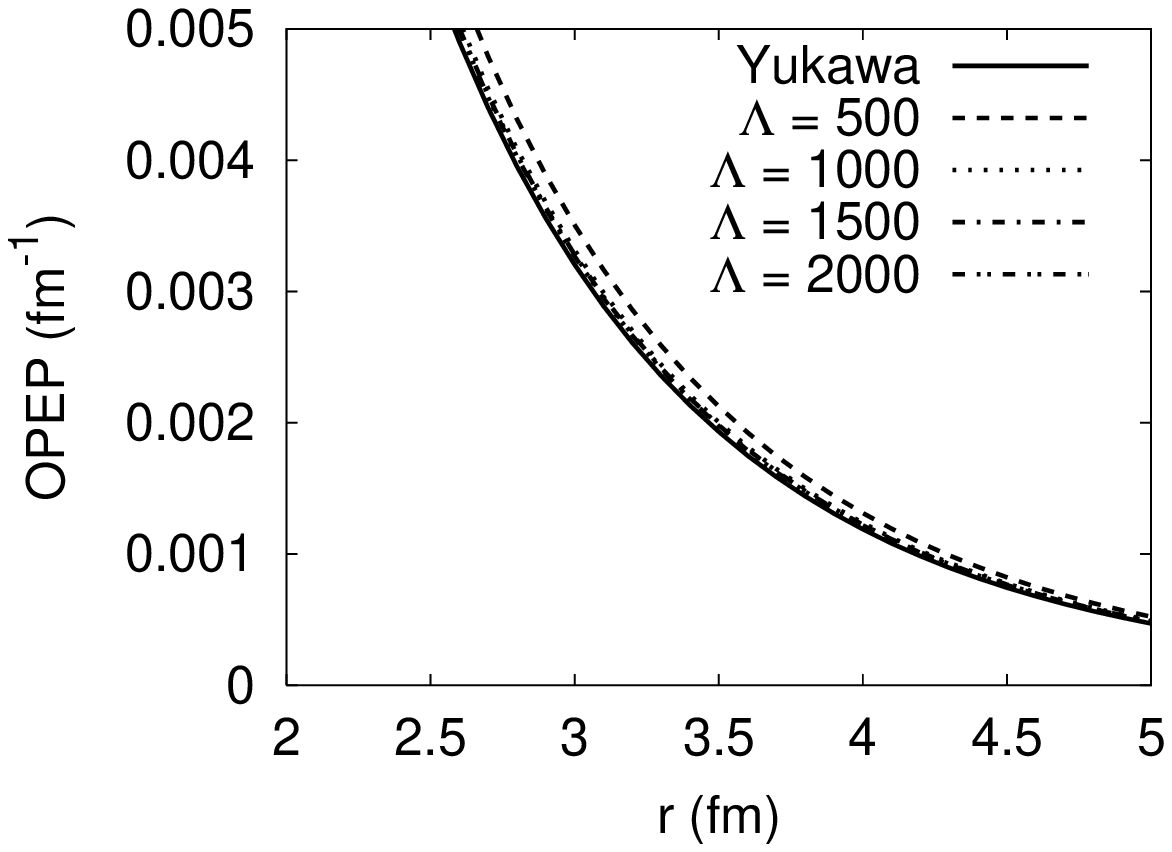, width=5cm} \\
\epsfig{file = 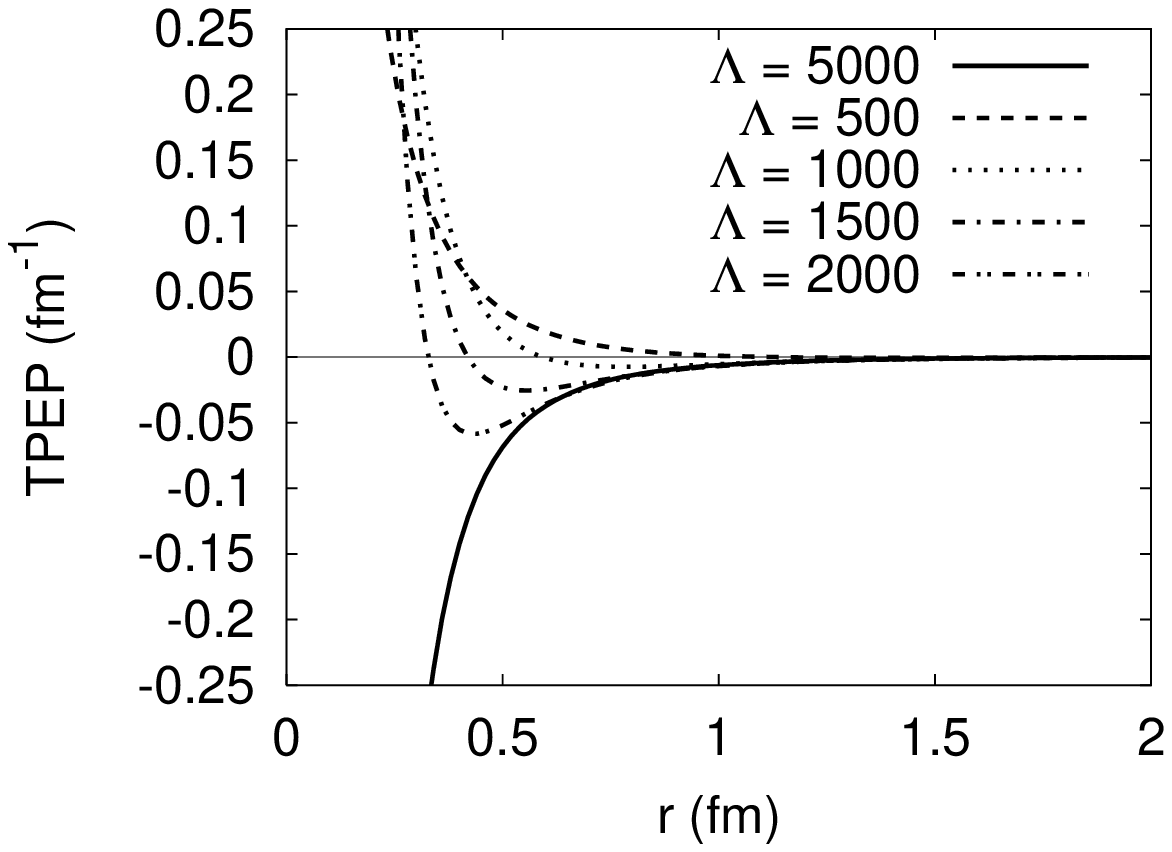, width=5.2cm}
\epsfig{file = 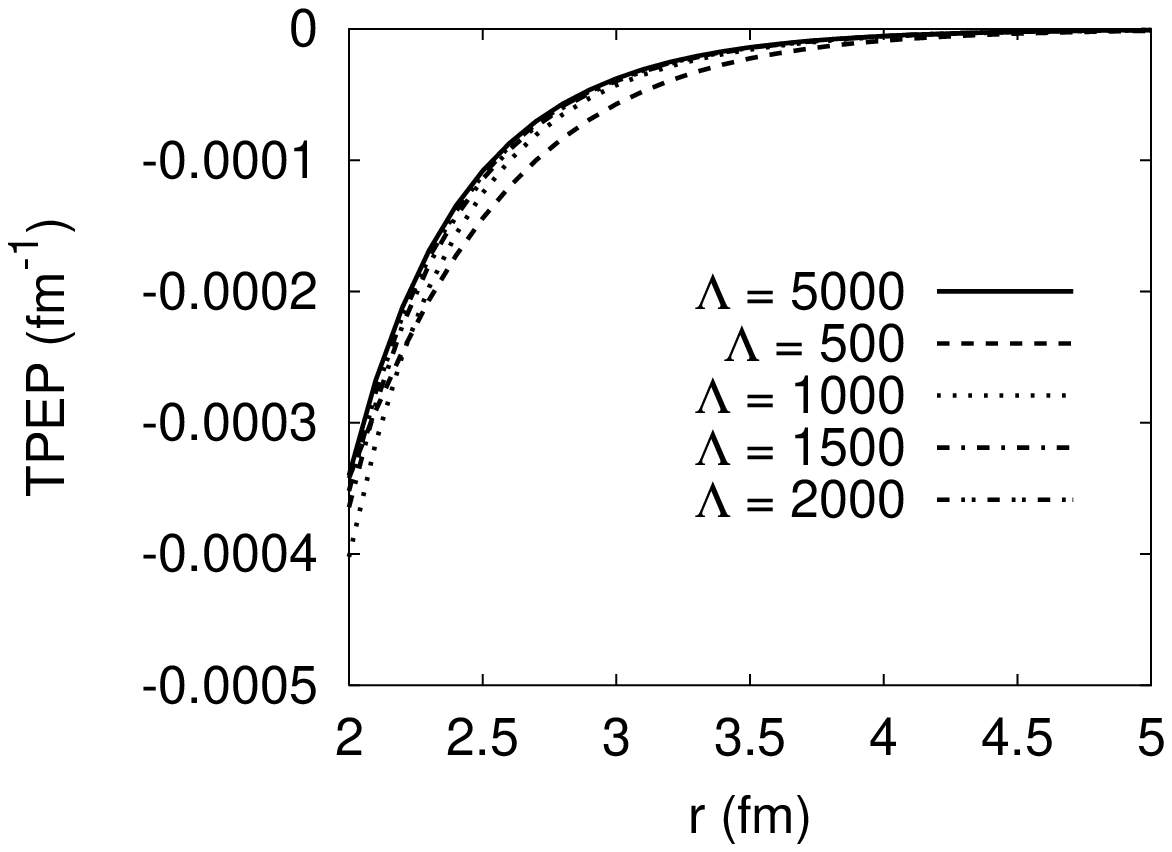, width=5.2cm}
\end{center}
\caption{$v^\Lambda_{1\pi}(r)$ (upper row) and 
$v^\Lambda_{2\pi}(r)$ (lower row) in the 
short-intermediate (left column) and long (right column) ranges.}
\label{fig:potential}
\end{figure*}

Fig.~\ref{fig:potential} shows 
$v^\Lambda_{1\pi}(r)$ (eq.~(\ref{eq:vr_ope})) and $v^\Lambda_{2\pi}(r)$
(eq.~(\ref{eq:vr_tpe})) in the range $0 < r < 5$ fm as functions of
cutoff $\Lambda$.
The curve denoted by `Yukawa' in OPEP corresponds to an infinite
cutoff value.
In the long range region (right panels), both OPEP and TPEP seldom
depend on $\Lambda$.
The magnitude of TPEP in the long range is smaller than OPEP by 
an order of 10, which confirms the dominance of OPEP 
in the long-range region.
In the short range, e.g. $r < 1$ fm, substantial cutoff dependence
is observed. The potential converges to the limiting case 
(infinite cutoff) with increasing $\Lambda$.
The change due to the cutoff value is simple in OPEP, but
TPEP shows more diversity in the dependence on the cutoff.
TPEP is a decreasing function at small $r$ but the sign of the
curvature changes at a certain value of $r$ and then it becomes increasing.
The value of $r$ at which the curvature of TPEP becomes zero is
sensitive to $\Lambda$, and it gives significant effect to $a_\gamma$.

\begin{table}
\caption{OPE and TPE contributions to
the asymmetry as functions of cutoff $\Lambda$.}
\label{tab:result}
\begin{center}
\begin{tabular}{crrrr}
\hline\noalign{\smallskip}
$\Lambda$ (MeV) & 500 & 1000 & 1500 & 2000 \\
\noalign{\smallskip}\hline\noalign{\smallskip}
$a_\gamma$ (OPE) & 
$-0.1074$ & $-0.1125$ & $-0.1126$ & $-0.1124$ \\
$a_\gamma$ (TPE) & 
$-0.0022$ & 0.0073 & 0.0117 & 0.0133 \\
\noalign{\smallskip}\hline
\end{tabular}
\end{center}
\end{table}

In Table~\ref{tab:result}, we show the numerical result of $a_\gamma$
defined in eq.~(\ref{eq:agamma}).
With OPEP, $a_\gamma$ is stable against the variation of
the cutoff value, and there is about 5\% fluctuation at most
with the cutoff values considered in the present work.
The TPEP contribution to $a_\gamma$, on the other hand,
varies significantly in magnitude and even sign change occurs.
From the behavior of $v^\Lambda_{1\pi}(r)$ and its contribution
to $a_\gamma$, it can be deduced that a decreasing function gives
negative contribution to $a_\gamma$.
With a larger $\Lambda$, the value of $r$ at which 
$\partial_r v^\Lambda_{2\pi}(r) = 0$ becomes smaller,
and the decreasing region of $v^\Lambda_{2\pi}(r)$ gets narrower.
This means that the negative contribution becomes smaller in magnitude
and a more positive contribution comes from TPEP.
This explains well the sign and magnitude behavior of $a_\gamma$ with
TPEP.

\section{Summary}

We have calculated the PV TPE 
$NN$ potential from the EFT, and applied it to the calculation
of PV asymmetry in $\vec{n} p \rightarrow d \gamma$
at threshold. In the intermediate and long range region,
TPEP is smaller than OPEP by an order of magnitude.
TPEP is comparable to OPEP for $r < 0.5$ fm.
Sizable cutoff dependence of the potential appears at $r \leq 1$ fm
for both OPEP and TPEP.
The asymmetry also shows non-negligible cutoff dependence.
For the asymmetry with OPEP, the dependence is relatively weak
and the result shows convergent behavior with larger cutoff values.
On the contrary, the asymmetry with TPEP is strongly dependent on
the cutoff value, even accompanied by a sign change, and the
convergence of the result is not clearly seen. 
As a result, the TPEP contribution to the asymmetry is about 
$-2\% \sim 12\%$ of the OPEP in the cutoff range $500 \sim 2000$ MeV.
Since 2000 MeV is a sufficiently large cutoff value,
the contribution of TPEP, though uncertain, 
lies in the range of (a few $\sim$ 10)\%
of OPEP. This is a significant correction to the OPEP result
compared to one-heavy-meson exchange, which is less than 1\%
of OPEP \cite{hyun_epja05}, but it is good to justify that
the perturbative expansion of EFT works well for the 
PV $NN$ potential.
The low energy constant terms at the NNLO, 
which are not considered in this work,
could give a similar amount of correction. 

%
%
%
%

\end{document}